\documentstyle[12pt,psfig]{article}

\def\FF{\hbox to 8.33887pt{\rm I\hskip-1.8pt F}}
\def\NN{\hbox to 9.3111pt{\rm I\hskip-1.8pt N}}
\def\PP{\hbox to 8.61664pt{\rm I\hskip-1.8pt P}}
\def\QQ{\rlap {\raise 0.4ex \hbox{$\scriptscriptstyle |$}}
{\hskip -4.5pt Q}}
\def\RR{\hbox to 9.1722pt{\rm I\hskip-1.8pt R}}
\def\ZZ{\hbox to 8.2222pt{\rm Z\hskip-4pt \rm Z}}

\newcommand{\resetequ}{\setcounter{equation}{0}}
\newcommand{\be}{\begin{equation}}
\newcommand{\ee}{\end{equation}}
\newcommand{\bqa}{\begin{eqnarray}}
\newcommand{\eqa}{\end{eqnarray}}
\newcommand{\ba}{\begin{array}}
\newcommand{\ea}{\end{array}}

\newcommand{\la}{\lambda}
\newcommand{\ph}{\phi}

\begin{document}

\centerline{\large \bf Constructive Field Theory and Applications:}
\centerline{\large \bf Perspectives and Open Problems}  
\vskip 2cm

\centerline{V. Rivasseau}
\centerline{Centre de Physique Th{\'e}orique, CNRS UPR 14}
\centerline{Ecole Polytechnique}
\centerline{91128 Palaiseau Cedex, FRANCE}

\vskip 1cm
\medskip
\noindent{\bf Abstract}

In this paper we review many interesting open problems in mathematical
physics which may be attacked with the help of tools from  constructive 
field theory. They could give work for future mathematical physicists 
trained with the constructive methods well within the 21st century.
\section{Introduction}
\resetequ

Constructive field theory started in the 70's as a program to study the 
existence and properties of non-trivial particular interacting field theories,
those with simple Lagrangians. Indeed it was not obvious at that time that
such structures (fulfilling a suitable set of axioms such as the Wightman 
axioms [Wi]) existed at all. In three decades, not only models of field 
theory, first superrenormalizable, then just renormalizable, have
been built and to some extent analyzed, but also the methods and techniques
developed in constructive field theory have been applied to a wide variety
of problems outside the initial scope of the program. 
Constructive techniques have been applied to equilibrium statistical 
mechanics, particularly to the study of critical phenomena, and to disordered 
systems. They have been introduced successfully in the analysis of many 
Fermions models, such as those of condensed matter. They have also
inspired and renewed studies in classical 
mechanics, and in time dependent problems, such as non-equilibrium phenomena.
It is no longer easy to draw the contours of this nebula. However the initial
group of people who pioneered constructive field theory in the early 70's,
together with a second and now a third generation of bright students, 
although working in very different domains nowadays,
still share in common a certain number of features. They are usually 
faithful to long-term programs, maybe even stubborn!  
Beyond the adjective ``constructive'', they share in common 
with the ``constructive'' trend in mathematics, advocated for instance
by Kronecker, a taste for explicit solutions, together with explicit bounds,
rathere than abstract existence theorems. In principle this means that
when translated into algorithms, and implemented on computers, the 
``constructive'' analysis of a physical model can lead to 
quantities computed with better precision and better controlled accuracy.

In this review for the special issue of JMP of year 2000,
we will be brief on the successes of the past and refer to the existing 
books\footnote{For classical references on 
constructive field theory, see [Er][Si][GJ1];
for reviews on constructive renormalization, and the problem of asymtotic
completeness, see respectively [R] and [Ia];
and for the most recent Proceedings on constructive field theory, see 
[CP].}. Instead we will focus on the open 
problems, conjectures and challenges that lie ahead in a subject that 
could now perhaps be called ``Constructive Physics'' rather than
``Constructive Field Theory'', and which remains characterized by the 
rigorous treatment of models issued from physics by hard analytic methods.
This paper does not contain any equation; its purpose is to entice the
reader to choose among the challenging problems just mentioned, and then
to go for the references, where
the formalisms for the corresponding problems are more precisely defined.
Finally we apologize for the fact that 
the list of open problems emphasized inevitably 
reflects our personal biases and interests. 

I thank D. Brydges and J. Imbrie for several discussions 
and for their hospitality at the University
of Virginia (Charlottesville, USA) where a large part of this 
paper was written; I thank also C. Kopper and V. Mastropietro for
several discussions.

\section{Constructive Field Theory}

This is the historical core of the theory, and in spite of some
spectacular successes, it remains largely a mine of open problems.

We recall that using the Euclidean functional integral approach, models
of non-trivial interacting field theories have been built over the past
thirty years, which satisfy Osterwalder-Schrader's axioms, hence in turn
have a continuation to Minkowski space that satisfy Wightman axioms
[Wi][OS][Z]. Such models are unfortunately yet restricted to space-time 
dimensions 2 or 3 but they include now both the first wave of 
superrenormalizable models, such as $P (\phi)_2$ [GJS1][GJ1][Si], 
$\phi^4_3$ [GJ2][Fe][FO][MS] or the Yukawa model in 2 and 3 
dimensions, as well as just renormalizable models such as the massive
Gross-Neveu model in two dimensions, or $GN_{2}$ [GK1,FMRS,DR1]. 
Most of these models have been built in the weak
coupling regime, using expansions such as the cluster and Mayer expansions;
the harder models require multiscale versions of these expansions, 
reshuffled according to the renormalization group philosophy.

In most cases the relationship of the non-perturbative construction
to the perturbative one has been elucidated, the non-perturbative Green's
functions being the Borel sum of the corresponding perturbative expansion
[EMS], [MS], [FMRS].

We identify and discuss several main areas for future progress:

\subsection{Asymptotic Freedom, Four dimensional Models}

By ``Coleman's theorem'', renormalizable
asymptotically free field theories in dimension
4 must involve non-Abelian gauge fields. 
However these fields lead to dreadful
infrared problems, e.g. confinement. Therefore no 
theory satisfying the flat infinite volume Wightman's axioms in dimension 4 
(the historic goal of constructive field theory)
has been constructed yet. However in a finite volume 
Balaban succeeded in proving ultraviolet stability of the effective 
action for non-Abelian lattice gauge theories (after an arbitrarily 
large number of  renormalization group steps) [Ba1]. We mention
also a less advanced attempt to construct this ultraviolet
limit in a particular non-standard gauge, using gauge symmetry breaking 
cutoffs [MRS]. This situation is not completely 
satisfying. We list among open problems:

\subsubsection{Non-linear sigma model}

Construct the ultraviolet limit of the
two dimensional O(N) non linear sigma model, which is a well behaved 
asymptotically free bosonic field theory 
(see e.g. [GK2] for construction of the hierarchical version of the model). 
It is quite irritating that we still do not have such a
construction: many experts in the field
tried it without success. The infrared mass generation 
has been obtained recently (see [K][IT]).

\subsubsection{Yang-Mills}

Construct the Yang Mills 4 correlation functions
in a finite volume and a standard gauge (such as the Landau gauge).
This presumably implies a front attack on the Gribov problem.

Simplify and rewrite Balaban's results on the lattice gauge theory. This is
no small task since references in [Ba1] total hundreds of pages\dots  

\subsubsection{$\ph_{4}^{4}$}

Elucidate the nature of $\ph_{4}^{4}$ renormalized pertubation series,
proving for instance that renormalons do exist (see [DFR]). Can one 
prove in full generality that its ultraviolet limit is trivial [A1][Fr][R]?

\subsubsection{Supersymmetric and Topological Field theory}

Develop constructive versions of the now-popular 
supersymmetric field theories, and 
topological field theories: develop a constructive understanding of
Montonen-Olive duality in $N=4$ SUSY Yang-Mills$_{4}$, and of Seiberg-Witten
duality for $N=2$ SUSY Yang-Mills$_{4}$.

Since these issues represent rather formidable challenges, it may be worth
to attack first some two dimensional problems [W1]: 
polynomial (hence superrenormalizable) $N=2$
supersymmetric field theories allow the construction
of interesting quantities sensitive to topological change [W2];
then the more difficult Wess-Zumino-Witten 
model at large parameter $k$, and the Calabi-Yau models, 
are field theories in which the field 
takes value in a non-trivial manifold as target space.

According to the point of view of Witten [W1], 
developing constructive theory
of functional field theoretic integrals for these models could attract more
mathematicians to field theory and may speed up the constructive programs
in other more traditional areas too.

\subsection{Strong Coupling or Low Temperature Results}

In the regime of strong coupling or low temperature, there are less
results. Contour expansions have been used to prove the existence
of the $\phi^4_2$ phase transition [GJS2]. Many results have been obtained for
models with an ultra-violet cutoff, i.e. models inspired by field theory but 
which are truly statistical mechanics models. For instance the phase transition
and non-perturbative mass generation has been proved for the $GN_{2}$
model with an ultraviolet cutoff [KMR], and continuous symmetry breaking (in 
dimensions greater or equal to 3) has
been studied with renormalization group techniques [Ba2]. 
An interesting challenge would be to glue this non-trivial low-temperature
analysis to the construction of the ultraviolet limit when it is possible.
Since the weak-coupling expansion for the ultraviolet limit is somewhat
in contradiction with the low-temperature expansion, this should be done
first for models with an other auxiliary small parameter, such as $N$-vector
models at large $N$, where the $1/N$ expansion can complete in the infrared
the small coupling expansion in the ultraviolet.

Therefore the first problems to attack in that direction could be:

\subsubsection{Constructive Dimensional Transmutation}\label{dimtrans}
Glue the ultraviolet analysis of the Gross-Neveu model [GK1][FMRS][DR1]
with the non-perturbative mass generation of the same model [KMR]
at large $N$, to obtain the first example of so called
dimensional transmutation.

One could also glue 
the ultraviolet construction of $\phi^4_3$ to the infrared continuous
symmetry breaking analysis of [Ba2] to control the 
large $N$-component $\phi^4_3$ model in the 
continuous symmetry broken phase without an ultraviolet cutoff.

\subsubsection{Constructive conformal field theory}

Develop rigorous links between conformal field theory in dimension two
and constructive field theory: along this line, the first significant
result should be to prove that the phase transition of $\phi^{4}_{2}$ 
is in the same universality class (i.e. has the same critical exponents) than
the Ising 2 phase transition. This could be later extended to 
$P (\phi_2)$ models with more vacua and Potts models.

More generally it should be nice to develop contact points between 
conformal field theory in two dimension [BPZ],
the theory of integrable systems, which relies more onto 
algebraic tools, and constructive theory which relies more 
on analysis. For instance there exist 
integrable lattice models of the ADE type which scale to
conformal field theories of the [BPZ] classification
at their critical point [Pa]; can we find a model which can be built with
constructive methods for every such conformal theory?

\subsection{Scattering, asymptotic completeness and\\
Minkowski space}

Develop phase space analysis and non-perturbative methods for field theory
that work directly in Minkowski space. This should lead to first proofs
of asymptotic completeness (see [Ia]) for quantum field theory models.
 
The easiest model in this direction may be the Gross-Neveu model
since it is ``purely perturbative'':
although it is just-renormalizable, hence has a worse ultraviolet power
counting than $\phi^4_2$, it
can be written purely as a reshuffled perturbation series [DR1],
so that in order to build it directly in Minkowski space
one does not need to develop a theory of functional
integration in Minkowski space, based on stationary phase analysis; 
it should be enough to simply develop a renormalization group analysis
around the mass shell hyperbola, which should resemble the renormalization
group around the Fermi surface of section \ref{condmat}.

\subsection{Other problems}

Complete in full detail 
the construction of the first non renormalizable field theory,
the Gross-Neveu model in three dimensions  and large number of components 
[dCFdVMS][dV]. Construct other models of this type, for instance
the corresponding regime of the sine-Gordon model.

Develop constructive field theory in curved space-time.

Make better contact with the $C^{*}$-algebra approach (sometimes
called axiomatic field theory). We refer to [BV] for a point of view on
the renormalization group in algebraic field theory.

\section{Equilibrium Statistical Mechanics}

In this category we already mentioned
the constructive study of continuous symmetry breaking [Ba2]
and dimensional transmutation in subsection \ref{dimtrans}.

\subsection{Coulomb gases}

After proofs of Debye screening and Kosterlitz Thouless (KT) 
phase transition, this area (together with the study of the sine-Gordon
and Thirring model) remains very active among constructivists.
For background in this subject we refer to [BM].
Here is a list of open problems for which we thank D. Brydges:

\subsubsection{} Find a direct proof of
convergence of the Mayer expansion for dipoles at 
low activity (which does not use a cluster expansion). 
The dipole-dipole interaction should be smoothed at short 
distance so that it is stable.

\subsubsection{}
Convergence of the Mayer expansion at low activity for the 
KT phase of the 2D Coulomb gas. This is harder than the previous problem,
and involves presumably an effective analysis of that gas in terms of 
multipoles.

\subsubsection{}
Prove exponential screening in the 2D Coulomb gas at not particularly
small temperatures (down to the KT transition?).

\subsubsection{}
Control the correlations at the KT transition.

\subsubsection{}
Are the transitions between $\beta_{KT}$ and $1/2 \beta_{KT}$ in the
D=2 Coulomb gas visible in any correlations?

\subsection{Disordered Systems}

The proof by J. Imbrie that the three dimensional
Random Field Ising model develops symmetry breaking at zero temperature
[Im] remains a beautiful example in which rigorous
constructive methods have solved
a controversial physical issue. Disordered systems are common in nature
(conductors or semi-conductors with structural defects or doping,
spin glasses, real glasses, granular or porous media, etc...). They pose
particularly challenging mathematical problems, and we review here only a
few of them.

\subsubsection{Anderson model of an electron in a random potential}

Here after the main results of [FS2] and followers on the localization 
regime at high disorder or out of the continuous spectrum of the free 
Hamiltonian, we feel that the main area for open problems is the 
weak-coupling phase. It is expected
that the Anderson 2 model $And_{2}$ is always localized but with exponentially
small localization length when the coupling constant tends to 0. A proof
of this statement through constructive field theory methods seems to require
first the proof of decay of a single averaged Green's function $<G_{\pm}>$
on a scale proportional to $\la^2$, 
the square of the coupling constant (this also controls
the density of states of the system). A constructive analysis is under way,
based on sector decomposition (as in [FMRT1]), a random matrix analogy,
and Ward identities [MPR1]. Then the real study
of localization involves the study of $<G_{+}G_{-}>$, and requires
a  resummation of leading ladders plus a study of the associated
``Goldstone mode''. Therefore the whole program is certainly as hard to
complete as the BCS2 program for interacting fermions defined in section IV
below.

In dimension 3 one expects the small coupling phase to be diffusive,
hence the system should undergo an ``Anderson Mott'' phase transition from
insulating to conducting at a certain critical coupling. To prove this
one should again first control the decay on a length scale of $\la^{-2}$ of
a single averaged Green's function:
this seems much harder than in dimension 2 essentially for the same
reason that BCS3 is much harder than BCS2: the random potential viewed
as a random matrix between angular directions is not of the usual
type (i.e. is not independent identically distributed, see [MPR2]).
After that difficulty has been solved, however, the task of controlling 
the square modulus of the Green's function $<G_{+}G^{-}>$ should be easier 
than in dimension 2, since we expect diffusion rather than localization.

\subsubsection{Constructive Study of Spin Glasses.}

This area is not familiar to me, but it contains certainly very 
challenging problems which do not often belong to 
the culture of main stream constructivists. To
``solve'' in a constructive sense models like the
Sherrington-Kirckpatrik model is certainly an ambitious
goal for the future. One should understand 
the correct notion of states for the model (in particular in connexion
with the ultrametric structure conjectured by the physicists).
It would be fascinating to also understand in a more precise
and constructive sense the replica symmetry breaking tool of Parisi.
Recently an explicit formula for the partition function at low temperature
has been obtained [Kou].

\subsection{Polymers}

Polymers and self-avoiding walks (SAW) 
are related to zero-component field theories
and have been often studied by constructive theorists. Among the established
results are the works [A2][BS][HS][L][IM] which explore the behavior
of these systems in 4 dimensions or more. Scaling dimensions of SAW 
with specific interactions in two dimensions can also be studied
rigorously through conformal invariance. See [D]
for a recent result in this area using quantum gravity methods.
 
Here is a list of open problems:

\subsubsection{} 
Polymers with partly attractive interactions: prove that they scale
to Brownian motion in $d>4$, at least if the interaction is stable
and small. Existence of transitions when the interaction is stable
but attractive?

\subsubsection{} 
$d=4$ Self avoiding walk: find new proofs that the end-to-end
distance has an exponent of 1/2 with log corrections.

\subsubsection{} 
Find new proofs that random walk in random environment scales to 
Brownian motion in $d>2$.

\subsubsection{} Prove anything at all 
about the expected end-to end distance of a self avoiding 
walk in $d<4$.  Is it greater than that of simple random
walk? Does it have an exponent?  If it does, is the exponent
different from 1/2?

\subsubsection{} 
Prove that the scaling limit of True Self Avoiding Walk in $d>2$ is Brownian
Motion.

\subsection{Interfaces, Wulff construction}

The constructive study of functional integrals associated to
interacting surfaces (Polyakov's functional integral) is much harder
than the ordinary random walk. The importance of these functional integrals
(for instance in string theory) nevertheless justify that constructivists
should get interested into them. 

\subsubsection{Wetting}
The study of interfaces is more advanced for solid-on solid models than for
real models such as the Ising model in the two phase regime.

An important open problem is to
construct the non-trivial renormalization group fixed point for a 
solid-on-solid model of an interface with two competing exponentials.
This should be doable at least in the regime where this fixed point is closed
to a Gaussian one, thanks to a small parameter in the rate of the two 
exponentials [BHL].

An other important problem is to give rigorous meaning to the Wulff
construction for such models [DM].

There are also perturbative results on the renormalization
surfaces interacting e.g. with a single impurity [DDG]
which one would like to connect to a constructive analysis.

\subsection{Non Equilibrium Statistical Mechanics}

The study of 
situations far from equilibrium made a big conceptual progress with
the introduction of the SRB steady states [SRB]. 
A typical recent rigorous result in this domain
is the fluctuation theorem of Gallavotti and Cohen [GC] on entropy
production. See [G1] for a discussion of this result.

We remark that the quantum non-equilibrium statistical mechanics
remain a widely open subject. An important long term goal
for constructive theory should be, after many body systems are
better understood and the main problems of the next section IV are solved, 
to develop the corresponding theory near
equilibrium, namley to put on a firm mathematical microscopic analysis
the Kubo formula and the Joule effect, and more generally transport theory.

\section{Condensed Matter}\label{condmat}

In the constructive theory of condensed matter, the main event of the
past was the adaptation of renormalization group techniques to models
with a Fermi surface [BG][FT1-2][FMRT1-5]. 

\subsection{Interacting fermions in 2 dimensions}
- In two dimensions there is a well-defined strategy which
should lead ultimately to the complete construction of the BCS2 model,
namely the control of the BCS phase at zero temperature [FMRT5]. There exists
already a control of the model until a scale where the coupling constant
becomes small but of order unity, which proves that any transition
temperature has to be exponentially small in the coupling 
[DR2-3]. Then the zone where the coupling 
constant is of order unity should be under control through some kind of 
$1/N$ expansion, where here $N$ is no longer an ad hoc parameter 
but is the effective number 
of angular directions on the Fermi surface at the scale considered
[FMRT2]; this 
expansion is not easy to write, and one may start with a simpler model
which has only quartic interaction at the BCS scale, like in [KMR]; then
one has to glue this analysis to the previous one, hence treat the 
corrections to the quartic effective action. Finally
one has to control the distance scales much longer that the BCS scale, where
the physics is governed by the infrared singularity of 
the Goldstone boson. Here the key tool should be a multiscale 
renormalization group analysis that relies on Ward identities [FMRT4] 
like in [Ba2]. This is a long and difficult program (even by 
the constructive standards!).

\subsection{Interacting fermions in 3 dimensions (BCS3)}
In dimension 3 the BCS program is less advanced. Although
perturbative power counting for the Fermi liquid is independent
of dimension, and the Goldstone boson problem is easier in 3+1 than 
in 2+1 dimensions, the initial regime (the equivalent of 
[FMRT1]-[DR2]) is harder to control for BCS3 because 
the momentum conservation laws 
are not as restrictive in 3 than in 2 dimensions: vertices can be non-planar,
or ``twisted'' in 3 dimensions [FMRT3]. 
The only rigorous result so far is that the radius of 
convergence of perturbation theory in a slice around the Fermi surface is 
independent of the distance of that slice to the singularity [MR]).
To find the analog of [FMRT1][DR2], namely that the sum of all ``convergent
contributions'' to the theory is analytic in the coupling constant 
remains in our opinion a major challenge of constructive theory.

\subsection{Bose-Einstein condensation}

Develop the theory of Bose-Einstein condensation. This can be viewed as a 
piece of the previous BCS program where the bosons are Cooper pairs, i.e. 
bound states of Fermions, or as an independent program if the bosons are
given from the start (see [Be]).
 
\subsection{Non-Spherical Surfaces; Hubbard Model}
 
\subsubsection{} Treat non-spherical surfaces. After the work
on the renormalization of convex surfaces [FKLT][FST], treat
surfaces with flat pieces
and/or singular points: the regular Hubbard model at half-filling on
a square lattice has both these features. Until now, this has to rely
for at least some part on numerical rather than analytical tools.

\subsubsection{} Develop a rigorous non-perturbative 
mean-field theory for condensed matter, i.e. develop the non-perturbative
version of the dynamical mean-field or $d=\infty$ limit of models such
as the Hubbard model: this dynamical mean-field model is really
a one dimensional theory with a self-consistent condition, 
but without an explicit action [GKKR].

\subsection{Quasi-periodic potentials, quasi-crystals}
 
Develop the mathematical theory of conduction in quasi-crystals.

In one dimension it is believed that fermions develop a Charge Density Wave
instability at small temperature with period equal to the inverse of the 
density. An interesting goal is to prove the
generation of such CDW in a system of interacting Fermions.
In this direction an expansion for interacting Fermions with an incommensurate
external potential satisfying a proper diophantine conditions 
was shown to converge in [Ma]; this is a first bridge on the gap between 
solid state physics and classical mechanics (the KAM theorem below),
since it amounts to solve a small denominator problem ``with loops''.
It would be nice to extend this bridge to other models, in particular in 
greater dimensions.

\section{Classical Mechanics}

Again this is an area I do not fell too competent to review and my remarks will
be brief. Contributions from constructive theorists have been devoted 
in particular to the area of the KAM theorem, where in particular the Italian
school around G. Gallavotti has developed the renormalization group
approach to the KAM theorem, but also to classical and quantum chaos, 
and to classical mechanics in random environment.

\subsection{KAM theory}

Invariant tori in Hamiltonian systems analytically close to integrable 
systems can be written as perturbative Lindstedt series in the perturbation
parameter. A direct proof of the convergence of such series, done by 
Eliasson [E], can be also obtained in the quantum field theory language,
using multiscale analysis as in the renormalization group,
and cancellations [G2]. (These cancellations 
can be interpreted as Ward identities
related to translation invariance [BGK1]).

An interesting open question is what happens to Lindstedt series in the 
non-analytic case. Moser, by using Nash theorem, proved that KAM tori
exist also in this case, with suitable conditions, but in general they are
not analytic. In [BGGM] analyticity was nevertheless proved for a class
of non-analytic pertubations by direct analysis of new
cancellations in the Lindstedt series. In more general cases where 
the summability of Lindstedt series may fail it is an open question to know if
some extended notion of summability, like Borel summability (quite frequent
in quantum field theory), may still hold.

Another set of problems concern Arnold's diffusion. In a 
priori unstable systems, a key quantity is the splitting which is the 
determinant of a certain matrix whose elements are series whose first
order is exponentially small, but the others are not. However the determinant
itself is exponentially small due to cancellations. Using Dyson equation
for classical mechanics, Arnold diffusion can be proved in certain a priori 
stable systems [GGM], but the same question is open in general a priori
stable systems such as those arising from celestial mechanics.

\subsection{Classical Chaos, Turbulence}

Of course the solution of Navier-Stokes equation and their scaling laws
remain a challenge, pretty much as it was at the beginning of the century. 
In the fully developed 
turbulence, experimentally observed
deviations from Kolmogorov's scaling of the velocity correlators
signal a non-Gaussian character of the velocity distributions at 
short distances, called intermittency.
Such intermittency, or deviations from Kolmogorov's scaling,
has been more or less understood in the particular case
of the passive advection of a scalar quantity 
(temperature, or density of a pollutant) by a random velocity field 
[GK3][BGK2]. However an explanation of the origin of intermittency 
in the general case of developed turbulence remains one of the main open 
problems of theoretical hydrodynamics.

\subsection{Quantum Chaos}

Here let us mention the results [CRR] on the Gutzwiller trace formula,
that one would like to extend to longer time evolution. A main challenge
is to put the heuristic connection between quantum chaos and the spectra 
of random matrices on a mathematically rigorous footing.

\subsection{Partial Differential Equations and Renormalization}

I would like to cite the work of Bricmont and Kupiainen on random walks
in a random environment [BK1], and more generally the application of 
renormalization group methods to partial differential equations [BK2].

The major open problems listed in [BK2] are the study of stability
of fronts in dissipative equations; extension of renormalization
group methods to hyperbolic equations; the study of invariant measures
for dynamical systems called 
Coupled Map Lattices [Ka], and of nonequilibrium ``phase transitions''
in which these invariant measures change as the coupling is varied.

\section{Improving Constructive Techniques}

In this section we would like to gather some list of mathematical
techniques which are quite general, so that they ought to be useful not only 
for a single problem but for many models in different branches of physics.

\subsection{Renormalization Group}

A central problem in constructive theory is to simplify
and further rationalize the various techniques which
allow to perform rigorous Renormalization Group computations.
The inductive version of the renormalization group itself has been better 
formalized by Brydges and coworkers [Br]; the multiscale phase space 
expansions which are some kind of expanded  
solution of the renormalization
group induction have been also recently formalized more explicitly [AR], 
and also recast using wavelets [Bat]. These efforts should be continued
if we want the rigorous approach to become part of the regular 
cursus of field theory. An open problem which could be
mentioned along these lines is to find an inductive rigorous constructive
renormalization which would be as simple as Polchinski's induction for
perturbative renormalization [P]: even for Fermions, this remains
an open problem [S2][DR1].

\subsection{Gluing together various expansions} 
Techniques to glue together different expansions or different
regimes of the renormalization group
(e.g small coupling/1/N coupling) should be developed. This is a
condition to treat many interesting models with ``non-perturbative''
phases. Somewhat like the geometric description of non-trivial manifolds
requires to glue several local charts together, the construction of 
non-trivial models with non-perturbative effects 
requires to develop
some experience in such gluing operations (see \ref{dimtrans}).

In a similar vein it is interesting to combine together several
expansion techniques which are usually treated separately.
For instance one can study the Many Body Models of section IV with
the additional complication of random or quasi-periodic environment
(see [Ma] for a one-dimensional example).

\subsection{Symmetries and Ward identities} 
Many difficult constructive problems involve symmetries which
are crucial to their understanding (gauge symmetries, supersymmetry,
replica symmetry). One would like to have
more general methods to quotient out or break
these symmetries, and develop
a more general theory of non-perturbative Ward identities.

\subsection{Non-integer dimensions} 
Non-integer dimensions is an interesting perturbative tool
(e.g. for the renormalization of non-Abelian gauge fields or 
for the $\epsilon$ expansion in statistical mechanics)
that has no constructive analog. One should understand why and
build the non-perturbative theory of functional integration
in non-integer dimensions. This is a long-term difficult goal,
perhaps related to non-commutative geometry, where ordinary space
is lost and the ordinary algebra of functions is replaced by
a non-commutative algebra.

\subsection{Random Matrices} 
Random matrices is a powerful tool for a wide range of physical problems,
from nuclear physics to quantum chaos, localization, 
quantized gravity and M-theory. The classical theory is the theory
of  independent identically distributed random matrices, and relates
them to orthogonal polynomials and integrable PDE's [M].
An important progress may come
from the understanding of random matrices with non-independent
coefficients. In the point of view of Voiculescu [V],
the Wigner law for independent identically distributed matrices model
is the non-commutative analog of Gaussian integration. Constructivists,
just as they developed the theory of non-Gaussian functional integration, 
may therefore try to develop a more general theory of random matrices, 
including in particular those with
constraints of geometric origin (see eg [MPR2] for an example).
This could presumably be very useful for the physics in spatial dimensions
higher than 2 (condensed matter, scattering, phase transitions).

\section{String Theory and Conclusion}
\medskip

When the constructive field theory program began in the 60's, field theory
was the prominent candidate for a fundamental theory of nature at the 
microscopic level (although it did not include quantization of gravity).
Today  the main stream of theoretical physics holds the view that field theory
is only an effective theory and that superstring or M-theory is the best 
candidate for a fundamental global theory of nature, a ``theory of 
everything''. Even if on a philosophical level the very existence of such a 
final theory is dubious, it is certainly a fascinating dream.
So in order to remain faithful to its initial quest, one could ask whether
constructivists should not join the efforts to find and build this TOE?

I would be tempted to adopt a rather cautious answer to this question,
namely ``Perhaps, but not yet''. There are three reasons for this cautious
attitude. 

- String theory or M-theory are mathematically very difficult: even the 
perturbative theory of superstring amplitudes contain enormous difficulties:
a proof of finiteness e.g. of the 10-dimensional $E_{8} \times E_{8}$ 
heterotic superstring amplitudes is a very difficult program in itself.

- The theory is in such a state of rapid evolution that it is not clear 
what should really be built. In the recent years, the different models had
a rather short life time before they were absorbed in a more 
general formalism. Under such circumstances, to launch a major constructive
effort could be premature, since the model might
be outdated well before the rigorous construction is completed.
 
- The theory has not yet received direct experimental confirmation. We
can at best hope for indirect hints, which may come in the next decades 
(spatial experiments such as those probing the background cosmic radiation, 
large cosmic rays detectors, new accelerators such as the LHC,
etc\dots may select along various cosmological or high energy scenarii, and
give indirect support to such or such models).

For these three reasons I do not think that time is ripe to launch today 
``constructive string theory'', as ``constructive field theory'' was launched
by A. Wightman and followers in the 60's.

To soften slightly these remarks, 
let me add that of course I consider string theory extremely important
for the future of mathematical physics. Indeed
string theory has not only been a very successful motivation to attract some
of the best minds to theoretical physics and to lead them to brilliant 
insights; it has also opened up a new interface
with mathematicians, mostly centered around geometry (differential, symplectic
and algebraic geometry, mirror symmetry, quantum cohomology, knot theory,
...). However this rapidly growing interface is very different from
the one opened in the past by constructive theory.
Algebra and geometry dominate over analysis, and there are no longer 
precise programs centered around axioms; 
but various pieces of the theory and various cross-consistent results
emerge progressively from this interaction between mathematicians
and theoretical physicists.
 
In conclusion, although at the present stage I would still rather personally
favor the applications of constructive field theory methods to well 
established physics, I would be happy, when some of the dust has settled, to 
see new generations of mathematical physicists attack in the
constructive spirit the problem of building rigorously
the high energy models that 
will emerge and survive in the coming century. 

\medskip
\noindent{\large{\bf References}}
\medskip
\vskip.1cm
%{\tenrm
{\small

\noindent [A1] M. Aizenman, {\em Geometric Analysis of $\phi^{4}$
fields and Ising models}, Comm. Math. Phys. {\bf 86}, 1 (1982). 
\vskip.1cm

\noindent [A2] M. Aizenman, Comm. Math. Phys. {\bf 97}, 91 (1985). 
\vskip.1cm

\noindent [AR] A. Abdesselam and V. Rivasseau, 
{\em An Explicit Large Versus Small Field Multiscale
Cluster Expansion}, Rev. Math. Phys. Vol. {\bf 9} No 2, 123
(1997). 
\vskip.1cm

\noindent [Ba1] T. Balaban,
Comm. Math. Phys. {\bf 95}, 17 and {\bf 96}, 223 (1984); {\bf 98}, 17, 
{\bf 99}, 75, {\bf 99}, 389 and {\bf 102},  277 (1985); {\bf 109},  
249 (1987); {\bf 116}, 1 and {\bf 119},  243 (1988); 
{\bf 122}, 175 and {\bf 122}, 355 (1989).
\vskip.1cm

\noindent [Ba2] T. Balaban, Comm. Math. Phys. {\bf 167} 103 (1995); {\bf 175},
607 (1996); {\bf 182}, 33 (1996); {\bf 182}, 675 (1997); 
{\bf 196}, 485 {\bf 198}, 1 and {\bf 198}, 493 (1998).
\vskip.1cm

\noindent [Bat] G. Battle, {\em Wavelets and Renormalization},
World Scientific, 1998.
\vskip.1cm

\noindent [Be] G. Benfatto, {\em Renormalization Group approach to Zero
Temperature Bose Condensation}, in [CP].
\vskip.1cm

\noindent [BF] D. Brydges and P. Federbush,
{\em Debye Screening}, Comm. Math. Phys. {\bf 73}, 197 (1980).
\vskip.1cm

\noindent [BG] G. Benfatto and G. Gallavotti,
{\em Perturbation theory of the Fermi surface in a quantum liquid.
A general quasi particle formalism and one dimensional systems}, 
Journ. Stat. Phys. {\bf 59} (1990) 541.
\vskip.1cm

\noindent [BGGM] F. Bonetto, G. Gallavotti, G. Gentile
and V. Mastropietro, {\em Quasi linear flows on tori and regularity
of their linearization},  Comm. Math. Phys. {\bf 192}, 707 (1998).
\vskip.1cm

\noindent [BGK1] J. Bricmont, K. Gawedzki and A. Kupiainen,
{\em Field Theory and KAM tori}, Comm. Math. Phys. {\bf 201}, 699 (1999).
\vskip.1cm

\noindent [BGK2] D. Bernard, K. Gawedzki and A. Kupiainen,
{\em Anomalous Scaling in the N-point functions of a Passive Scalar},
Phys. Rev. {\bf E54}, 2564 (1996).
\vskip.1cm

\noindent [BK1] J. Bricmont and A. Kupiainen,
{\em Random walks in asymmetric random environments}, 
Comm. Math. Phys. {\bf 142}, 345 (1991).
\vskip.1cm

\noindent [BK2] J. Bricmont and A. Kupiainen,
{\em Renormalizing Partial Differential Equations}, 
in [CP].
\vskip.1cm

\noindent [BHL] E. Br{\'e}zin, B. I. Halperin and S. Leibler,
Phys. Rev. Lett. {\bf 50}, 1387 (1983).
\vskip.1cm

\noindent [BM] D. Brydges and Ph. Martin, {\em Coulomb Systems at low density},
to appear in Journ. Stat. Phys.
\vskip.1cm

\noindent [BPZ] Belavin, Polyakov and Zamolodchikov, Nucl. Phys. {\bf B241}, 
333 (1984)
\vskip.1cm

\noindent [Br] D. Brydges, {\em Weak perturbations of the Massless Gaussian
Measure}, in [CP].
\vskip.1cm

\noindent [BS] D. Brydges and T. Spencer, 
Comm. Math. Phys. {\bf 97}, 125 (1985).
\vskip.1cm

\noindent [BV] D. Buchholz and R. Verch, 
Rev. Math. Phys. {\bf 7}, 1195 (1995)
\vskip.1cm

\noindent [CP] Constructive Physics, ed by
V. Rivasseau, Lecture Notes in Physics {\bf 446}, Springer Verlag, 1995.
\vskip.1cm

\noindent [CRR] M. Combescure, J. Ralston and D. Robert, {\em A proof of the 
Gutzwiller semi-classical trace formula using coherent states decomposition},
Comm. Math. Phys. {\bf 202}, 463 (1999)
\vskip.1cm

\noindent [D] B. Duplantier, {\em Random Walks, Polymers, Percolation
and Quantum Gravity in Two Dimensions}, Physica A {\bf 263}, 452 (1999).
\vskip.1cm

\noindent [dCFdVMS] de Calan, C., Faria da Veiga, P.A., Magnen, 
J., S{\'e}n{\'e}or, R.: {\em Constructing the Three Dimensional Gross-Neveu 
Model with a Large Number of Flavor Components}, Phys. Rev. Lett. {\bf 66}, 
3233-3236 (1991).
\vskip.1cm

\noindent [DDG] F. David, B. Duplantier and E. Guitter, 
{\em Renormalization Theory for the Self-Avoiding Polymerized Membranes},
cond-mat/9702136.
\vskip.1cm

\noindent [DFR] F. David, J. Feldman and V. Rivasseau
{\em On the large order behavior of $\phi^{4}_{4}$}, 
Comm. Math. Phys. {\bf 116}, 215 (1988).
\vskip.1cm

\noindent [DM] F. Dunlop and J. Magnen, {\em A Wulff Shape from 
Constructive Field
Theory}, in Mathematical Results in Statistical Mechanics, World Scientific,
1998.

\vskip.1cm
\noindent [DR1] M. Disertori and V. Rivasseau,
{\em Continuous Constructive Fermionic Renormalization}, hep-th/9802145,
to appear in Annales Henri Poincar{\'e}.
\vskip.1cm

\noindent [DR2] M. Disertori and V. Rivasseau, {\em Interacting Fermi liquid 
in two dimensions at finite temperature, Part I: Convergent Attributions},
cond-mat/9907130.
\vskip.1cm

\noindent [DR3] M. Disertori and V. Rivasseau, {\em Interacting Fermi liquid 
in two dimensions at finite temperature, Part II: Renormalization},
cond-mat/9907131.
\vskip.1cm

\noindent [dV] P. Faria da Veiga, 
{\em Construction de Mod{\`e}les Non Perturbativement Renormalisables 
en Th{\'e}orie Quantique des Champs}, Th{\`e}se, Universit{\'e} de Paris XI, 1992; 
\vskip.1cm

\noindent [E] L.H. Eliasson, 
{\em Absolutely convergent series expansions for quasi-periodic motions}, 
Math. Phys. Elec. Journal, {\bf 2}, 1996. 
\vskip.1cm

\noindent [EMS] J.P. Eckmann, J. Magnen and R. S{\'e}n{\'e}or, 
{\em Decay properties and  
Borel summability for the Schwinger functions in $P(\phi)_{2}$ theories}, 
Comm. Math. Phys. {\bf 39}, 251 (1975).
\vskip.1cm

\noindent[Er] Constructive Quantum field theory, Proceedings of the 1973  
Erice Summer School, ed. by G. Velo and A. Wightman,   
Lecture Notes in Physics, Vol. {\bf 25}, Springer 1973. 
\vskip.1cm

\noindent [Fe] J. Feldman, {\em The $\lambda\phi^{4}_{3}$ field theory 
in a finite volume}, Comm. Math. Phys. {\bf 37}, 93 (1974).
\vskip .1cm

\noindent [Fr] J. Fr{\"o}hlich, {\em On the triviality of $\la\ph^{4}_{d}$
theories and the approach to the critical point in $d\ge 4$ dimensions},
Nucl. Phys. {\bf B200} (FS4), 281 (1982).
\vskip .1cm

\noindent [FKLT] J. Feldman, H. Kn{\"o}rrer, D. Lehmann and E. Trubowitz, 
{\em Fermi Liquids in Two Space Time Dimensions}, in [CP].
\vskip.1cm

\noindent [FMRS1] J. Feldman, J. Magnen, V. Rivasseau and R.  
S{\'e}n{\'e}or, {\em A renormalizable field theory: 
the massive Gross-Neveu model in two  
dimensions}, Comm. Math. Phys. {\bf 103}, 67 (1986). 
\vskip.1cm
 
\noindent [FMRT1] J. Feldman, J. Magnen, V. Rivasseau and E. Trubowitz,
{\em An infinite Volume Expansion for Many Fermion Green's Functions},
Helv. Phys. Acta {\bf 65} (1992) 679.
\vskip.1cm

\noindent [FMRT2] J. Feldman, J. Magnen, V. Rivasseau and E. Trubowitz,
{\em An Intrinsic 1/N Expansion for Many Fermion System}, Europhys. Letters 
{\bf 24}, 437 (1993).
\vskip.1cm

\noindent [FMRT3] J. Feldman, J. Magnen, V. Rivasseau and E. Trubowitz,
{\em Two dimensional Many Fermion Systems as Vector Models}, Europhys. Letters 
{\bf 24}, 521 (1993).
\vskip.1cm

\noindent [FMRT4] J. Feldman, J. Magnen, V. Rivasseau and E. Trubowitz,
{\em Ward Identities and a
Perturbative Analysis of a U(1) Goldstone Boson in a Many Fermion System}, 
Helv. Phys. Acta {\bf 66}, 498 (1993).
\vskip.1cm

\noindent [FMRT5] J. Feldman, J. Magnen, V. Rivasseau and E. Trubowitz,
A R{\em igorous Analysis of the Superconducting Phase of an 
Electron-Phonon System}, 
Proceedings de l'Ecole des Houches 1994 (F. David, P. Ginsparg eds.

\noindent [FO] J. Feldman and K. Osterwalder, {\em The Wightman axioms and the 
mass gap for weakly coupled $\phi^{4}_{3}$ quantum field theories}, 
Ann. Phys. {\bf 97}, 80 (1976).
\vskip.1cm

\noindent [FS1] J. Fr{\"o}hlich and T. Spencer, {\em The Kosterlitz-Thouless
Transition in Two-Dimensional Abelian Spin Systems and the
Coulomb Gas}, Comm. Math. Phys. {\bf 81}, 527 (1981).
\vskip.1cm

\noindent [FS2] J. Fr{\"o}hlich and T. Spencer, {\em Absence of diffusion in the 
Anderson tight binding model for large disorder or low energy},
Comm. Math. Phys. {\bf 88}, 151 (1983).
\vskip.1cm

\noindent [FST] J. Feldman, M. Salmhofer and E. Trubowitz, 
{\em Perturbation Theory 
around Non-nested Fermi Surfaces II.  Regularity of the Moving Fermi 
Surface, RPA Contributions}, 
Comm. Pure. Appl. Math. {\bf 51} (1998) 1133;
{\em Regularity of the Moving Fermi Surface, The Full Selfenergy},
to appear in Comm. Pure. Appl. Math.
\vskip.1cm

\noindent [FT1]  J. Feldman and E. Trubowitz, 
{\em Perturbation theory for Many Fermion Systems}, Helv. Phys. Acta {\bf 63}
(1991) 156.
\vskip.1cm

\noindent [FT2]  J. Feldman and E. Trubowitz, {\em 
The flow of an Electron-Phonon
System to the Superconducting State}, Helv. Phys. Acta {\bf 64}
(1991) 213.
\vskip.1cm

\noindent [G1] G. Gallavotti, {\em A local Fluctuation Theorem},
Physica A {\bf 263}, 39 (1999).
\vskip.1cm

\noindent [G2] G. Gallavotti, {\em Twistless KAM tori, quasi flat homoclinic
intersections, and other cancellations in the perturbation series of certain
completely integrable Hamiltonian systems. A review}, Reviews on Mathematical
Physics {\bf 6}, 343 (1994).
\vskip.1cm

\noindent [GC] G. Gallavotti and E.G.D. Cohen,
{\em Dynamical ensembles in nonequilibrium statistical mechanics},
Phys. Rev. Lett. {\bf 74}, 2694 (1995); {\em Dynamical ensembles in 
stationary states},
Journ. Stat. Phys. {\bf 80}, 931 (1995).
\vskip.1cm

\noindent [GGM] G. Gallavotti, G. Gentile and V. Mastropietro,
{\em Hamilton-Jacobi equation and existence of heteroclinic chains in three
time scales systems} (to appear on Nonlinearity).
\vskip.1cm

\noindent [GJ1] J. Glimm and A. Jaffe, ``Quantum Physics. A functional 
integral point of view", Mc Graw and Hill, New York, 1981
\vskip .1cm
 
\noindent [GJ2] J. Glimm and A. Jaffe, {\em Positivity of the $\phi^{4}_{3}$  
Hamiltonian}, Fortschr. Phys. {\bf 21}, 327 (1973). 
\vskip .1cm

\noindent [GJS1] J. Glimm, A. Jaffe and T. Spencer, 
{\em The Wightman Axioms and Particle Structure in the $P (\phi)_{2}$ Quantum
Field Model}, Ann. Math. {\bf 100}, 585 (1974).
\vskip .1cm

\noindent [GJS2] J. Glimm, A. Jaffe and T. Spencer, 
{\em Phase transitions for $\phi^4_{2}$ Quantum Fields}, Comm. Math. Phys. 
{\bf 45}, 203 (1975).
\vskip .1cm

\noindent [GK1] K. Gawedzki and A. Kupiainen, {\em Gross-Neveu model through  
convergent perturbation expansions}, Comm. Math. Phys. {\bf 102}, 1 (1985). 
\vskip .1cm 

\noindent [GK2] K. Gawedzki and A. Kupiainen, {\em Comtinuum limit of the 
hierarchical $O(N)$ nonlinear $\sigma$ model}, Comm. Math. Phys. 
{\bf 106}, 533 (1986). 
\vskip .1cm 

\noindent [GK3] K. Gawedzki and A. Kupiainen, {\em Anomalous Scaling of the 
Passive Scalar}, Phys. Rev. Lett. {\bf 75}, 3834 (1995).
\vskip .1cm 

\noindent[GKKR] A. Georges, G. Kotliar, W. Krauth and M. Rozenberg,
{\em The Local Impurity Self-Consistent Approximation to Strongly Correlated
Fermion Systems and the Limit of Infinite Dimensions},
Rev. Mod. Physics, {\bf 68}, 13 (1996).
\vskip .1cm
 
\noindent [HS] T. Hara and G. Slade, Comm. Math. Phys. {\bf 147}, 101 (1992).
\vskip .1cm 

\noindent [Ia] D. Iagolnitzer, Scattering in Quantum Field Theories,
Princeton University Press, 1992.
\vskip .1cm 

\noindent [Im] J. Imbrie, {\em The Ground State of the 
Three Dimensional Random Field Ising Model},
Comm. Math. Phys. {\bf 98}, 145 (1985).
\vskip .1cm 

\noindent [IM] D. Iagolnitzer and J. Magnen, {\em Polymers in a Weak
Random Potential in Dimension 4: Rigorous Renormalization Group Analysis},
Comm. Math. Phys. {\bf 162}, 85 (1994).
\vskip .1cm 

\noindent [IT] K.R. Ito and H. Tamura,
{\em $N$-dependence of upper bounds of critical temperatures
of 2d O(N) spin models}, Comm. Math. Phys. {\bf 202}, 127 (1999).
\vskip .1cm 

\noindent [K] C. Kopper, {\em Mass Generation in the large $N$ non-linear 
sigma Model}, Comm. Math. Phys. {\bf 202}, 89 (1999).
\vskip .1cm 

\noindent [Ka] K. Kaneko, Theory and Applications of Coupled Map Lattices,
Wiley, (1993).
\vskip .1cm 

\noindent [KMR] C. Kopper, J. Magnen and V. Rivasseau, 
{\em Mass Generation in the Large N Gross-Neveu Model},
Comm. Math. Phys. {\bf 169}, 121 (1995)
\vskip .1cm 

\noindent [Kou] F. Koukiou, {\em Large deviations for the mean behaviour
of the Sherrington-Kirkpatrick spin glass 
model}, to appear in Comm. Math. Phys. (1999).
\vskip .1cm

\noindent [M] M. L. Mehta, Random Matrices and the Statistical Theory of 
Energy Levels, Academic Press 1967.
\vskip.1cm

\noindent [Ma] V. Mastropietro, {\em Small Denominators and Anomalous 
Behaviour in the Incommensurate Hubbard-Holstein Model}, Comm. Math. 
Phys. {\bf 201}, 81 (1999).
\vskip.1cm

\noindent [MPR1] J. Magnen, G. Poirot and V. Rivasseau, 
{\em Ward type Identities 
for the 2d Anderson Model at weak Disorder},
Journ. Stat. Phys. {\bf 93}, 331 (1998)
\vskip.1cm

\noindent [MPR2] J. Magnen, G. Poirot and V. Rivasseau,
{\em The Anderson Model as a Matrix Model}, 
Nucl. Phys. B (Proc. Suppl.) {\bf 58}, 149 (1997).
\vskip.1cm

\noindent [MR] J. Magnen and V. Rivasseau, {\em A Single 
Scale Infinite Volume Expansion for
Three Dimensional Many Fermion Green's Functions},
Math.  Phys. Electronic  Journal, {\bf 1},  1995.
\vskip.1cm

\noindent [MS] J. Magnen and R. S{\'e}n{\'e}or, {\em Phase space cell expansion 
and Borel summability for the Euclidean $\phi^{4}_{3}$ theory}, Comm Math. 
Phys. {\bf 56}, 237  (1977).
\vskip.1cm

\noindent [MRS] J. Magnen, V. Rivasseau and R. S{\'e}n{\'e}or,
{\em Construction of $YM_{4}$ with an infrared cutoff}, 
Commun. Math. Phys. {\bf 155}, 325 (1993).
\vskip.1cm

\noindent [P] J. Polchinski, Nucl Phys B {\bf 231}, 269 (1984)
\vskip.1cm

\noindent [Pa] V. Pasquier, {\em Two dimensional critical
systems labelled by Dynkin diagrams}, Nucl. Phys. {\bf B285}, [FS19]
162 (1986)
\vskip.1cm

\noindent [OS] K. Osterwalder and R. Schrader, {\em Axioms 
for Euclidean Green's  
functions}, Comm. Math. Phys. {\bf 31}, 83 (1973).
\vskip.1cm

\noindent [R] V. Rivasseau, From perturbative to constructive renormalization,
Princeton University Press, 1991.
\vskip.1cm

\noindent [Si] B. Simon, The $P(\phi)_{2}$ (Euclidean) Quantum field theory,  
Princeton University Press, 1974. 
\vskip .1cm

\noindent [S1] M. Salmhofer, {\em Improved Power Counting and Fermi Surface
Renormalization}, Rev. Math. Phys. {\bf 10}, 553 (1998).
\vskip.1cm

\noindent [S2] M. Salmhofer,
{\em Continuous renormalization for Fermions and Fermi liquid theory},
Commun. Math. Phys.{\bf 194}, 249 (1998).
\vskip.1cm

\noindent [SRB] Y. Sinai, {\em Gibbsian measures in ergodic theory},
Uspekhi Mat. Nauk {\bf 27 no 4}, 21 (1972); R. Bowen and D. Ruelle,
{\em The ergodic theory of Axiom A flows}, Invent. Math.
{\bf  29}, 181 (1975); D. Ruelle,
{\em A measure associated with Axiom A attractors}, Am. J. Math. {\bf 98}, 
619 (1976).
\vskip.1cm

\noindent  [V] D. Voiculescu, {\em Limit Laws for random Matrices and Free 
Products}, Invent. Math. {\bf 104}, 201 (1991). 
\vskip.1cm

\noindent [W1] E. Witten, {\em Some Questions for Constructive Field 
Theorists}, in [CP].
\vskip.1cm

\noindent [W2] E. Witten, {\em Phases of $N=2$ Models in two Dimensions}, 
Nucl. Phys. {\bf B403}, 159, (1993)
\vskip.1cm

\noindent [Wi]
A. Wightman, {\em Quantum field theory in terms of vacuum expectation 
values}, Phys. Rev. {\bf 101}, 860 (1956)
\vskip.1cm

\noindent [Z] Y. Zinoviev, {\em Equivalence of Euclidean and Wightman Field
Theories}, in [CP].

}
\end{document}